\documentclass[a4paper,fleqn, usenatbib]{mnras}

\usepackage[T1]{fontenc}
\usepackage{ae, aecompl}

\usepackage{graphicx}	
\usepackage{amsmath}	
\usepackage{amssymb}	
\usepackage{deluxetable}
\usepackage{pdflscape}
\usepackage{threeparttable}

\title[SEDs of EMP star-forming regions]{Spatially resolved dust emission of extremely metal poor galaxies\thanks{{\it Herschel} is an ESA space observatory with science instruments provided by European-led Principal Investigator consortia and with important participation from NASA.}}

\author[Luwenjia Zhou]{Luwenjia Zhou$^{1,2}$, Yong Shi$^{1, 2, 3}$,
Taino Diaz-Santos$^{4}$,
  Lee Armus$^{5}$,
  George Helou$^{5}$, 
  \newauthor
   Sabrina Stierwalt$^{6}$, and 
  Aigen Li$^{7}$
\\
$^{1}$School of Astronomy and Space Science, Nanjing University, Nanjing 210093, China\\
$^{2}$Key Laboratory of Modern Astronomy and Astrophysics (Nanjing University), Ministry of Education, Nanjing 210093, China\\
$^{3}$Collaborative Innovation Center of Modern Astronomy and Space Exploration, Nanjing 210093, China\\
$^{4}$Nu ́cleo de Astronom ́ıa de la Facultad de Ingenier ́ıa, Universidad Diego Portales, Av. Ej ́ercito Libertador 441, Santiago, Chile\\
$^{5}$Infrared Processing and Analysis Center, California Institute of Technology, 1200 E. California Boulevard, Pasadena, CA 91125, USA\\
$^{6}$Department of Astronomy, University of Virginia, P.O. Box 400325, Charlottesville, VA 22904, USA\\
$^{7}$Department of Physics and Astronomy, University of Missouri, Columbia, MO 65211, USA
}

\date{Accepted XXX. Received YYY; in original form ZZZ}

\pubyear{2015}

\begin{document}
\label{firstpage}
\pagerange{\pageref{firstpage}--\pageref{lastpage}}
\maketitle

\begin{abstract}
  We  present   infrared (IR)  spectral  energy  distributions   (SEDs)  of
  individual star-forming  regions in four extremely  metal poor (EMP)
  galaxies with metallicity $Z$  $\lesssim$ $Z_{\odot}$/10 as observed
  by the {\it  Herschel} Space Observatory.  With  the good wavelength
  coverage of the SED, it is found that these EMP star-forming regions
  show distinct SED shapes as  compared to those of  grand design Spirals
  and  higher metallicity  dwarfs: they  have on  average much  higher
  $f_{70{\mu}m}/f_{160{\mu}m}$ ratios at a   given
  $f_{160{\mu}m}/f_{250{\mu}m}$  ratio;  single  modified  black-body
  (MBB) fittings  to the  SED at  $\lambda \ge$  100 $\mu$m  still
  reveal higher dust temperatures  and lower emissivity indices compared 
  to that of Spirals, while two MBB  fittings  to  the full  SED  with a  fixed
  emissivity index ($\beta$ = 2) show  that even at 100 $\mu$m about
  half of the emission  comes from warm (50 K) dust, in contrast to the 
  cold ($\sim$20 K) dust component.  
  Our spatially  resolved images  further reveal  that the
  far-IR      colors      including      $f_{70{\mu}m}/f_{160{\mu}m}$,
  $f_{160{\mu}m}/f_{250{\mu}m}$ and  $f_{250{\mu}m}/f_{350{\mu}m}$ are
  all related  to the surface  densities of young stars  as traced by far-UV, 24
  $\mu$m  and   SFRs,  but   not  to   the  stellar   mass  surface
  densities. This suggests that the  dust emitting at wavelengths from
  70 $\mu$m  to  350 $\mu$m is  primarily heated by radiation
  from young stars.

\end{abstract}




\section{Introduction}\label{introduction}

Stars born within  primordial galaxies in  the early  universe form out  of gas
with  no  or little  metals.   These  distant galaxies  are,  however,
difficult to  detect. Nearby  extremely metal poor  (EMP) star-forming
galaxies   are  chemically   unevolved,  and   thus  offer   important
astrophysical laboratories for our  understandings of the interstellar
medium  (ISM)  and  star  formation in  the low  metallicity  environments
\citep{Kunth00, Remy13, Shi14,  Shi15}. Dust grains play  a vital role
in galaxy formation and evolution.  Characterizing the dust properties
by  investigating  the  infrared (IR) emission  is  a   powerful  way  to
understanding   the    ISM   and    evolution   of    dwarf   galaxies
\citep{Feldmann15}.

The integrated dust  properties of dwarfs have  been investigated with
$Spitzer$ Space  Telescope.  It  is found  that aromatic  features are
absent in the metal  poor galaxies.  \citet{Engelbracht05} pointed out
that  the ratio  f$_{8\mu m}$/f$_{24\mu  m}$ depends  strongly on  the
metallicity  (where f$_{\lambda}$  is the  flux density  at wavelength
$\lambda$), with significantly lower values (mean ratio 0.08 $\pm$
  0.04) for  galaxies below one-third of the  Solar metallicity
  than  those   at  higher   metallicities  (mean  ratio   0.70  $\pm$
  0.53).\footnote{\citet{Engelbracht05} adopted Solar metallicity
  as 12  + log(O/H)  = 8.7\citep{Allende01}.  Here  in this  paper, we
  define Solar  metallicity to be 8.7  as well. Thus one-third  of the
  Solar  metallicity is  around 8.2.}   \citet{Draine07} confirmed  no
 polycyclic    aromatic   hydrocarbons   (PAH)   emission   in
low-metallicity galaxies, as well  as other works \citep[e.g.][]{Wu06,
  Rosenberg08}.    The   HI-to-dust   mass   ratio   as   studied   by
\citet{Engelbracht08}  is  shown  to   increase  with  the  decreasing
metallicity to 12+log(O/H)$\thicksim$8,  as $\thicksim$Z$^{-2.5}$
, and  flattens out at  lower oxygen  abundances.  They also  found an
anti-correlation between  dust temperature  \footnote{ Temperatures
  are  derived  from modified  blackbody,  with  the emissivity  index
  $\beta$    fixed    to   be    2.}     and    metallicity   up    to
12+log(O/H)$\thicksim$8,  from  T$\thicksim$23 K  near the  Solar
  metallicity  to  T$\thicksim$40 K  at  lower  metallicity,  but  a
positive relation at lower metallicities.

The {\it  Herschel} Space  Observatory further extends  the wavelength
coverage  into  the  far-IR   and  sub-millimeter  wavelengths,  which
combined with the {\it  Spitzer} photometry significantly improves the
measurement of  the dust properties.   The total gas-to-dust  ratio as
revealed by {\it Herschel}  increases the decreasing metallicity.
  The  slope  of   the  relationship  is  -1   at  higher  metallicity
  (12+log(O/H) > 8) and becomes even steeper at the lower metallicity
end \citep{Remy14}.  It  is also found that the metal  content may not
be the only factor affecting the dust properties of dwarfs.
 For  example, the dust  to stellar mass ratio  of IZw 18  and SBS
  0335-052, which have similar metallicities to the other EMP galaxies
  in  Table~\ref{tab:properties},  are  very different  (10$^{-6}$  to
  10$^{-5}$, and 10$^{-3}$ to 10$^{-2}$, respectively) as estimated by
  \citet{Fisher14}  and \citet{Hunt14},  respectively.  The  ratio for
  IZw 18 is  extremely low, while that for SBS  0335-052 is comparable
  to that found for normal Spiral galaxies \citep{Hunt14}.

While  significant  progresses have  been  made  in understanding  the
integrated  dust  properties  of   EMP  galaxies,  spatially  resolved 
investigations of  dust properties of  these galaxies are  still rare,
and thus the  dependence of dust properties on the  local condition is
largely unconstrained in  the EMP galaxies.  In this study,  we aim to
benefit  from the  high  spatial resolution  of  the $Herschel$  Space
Observatory  to  investigate  the   IR  emission  of  individual
star-forming regions of  four EMP galaxies, with focus on  the
IR SED shapes and their relations to the local conditions.

\section{Sample, Observations And Data Analysis}\label{sec_data}

\subsection{The Sample}

Our  sample of   EMP galaxies  contains four  objects
including Sextans  A, ESO  146-G14, DDO  68 and  Holmberg II  (Ho II).
Sextans A is a dwarf irregular  at 1.4 Mpc with 12+log(O/H)=7.49 based
on the  direct method  \citep{Kniazev05}.  ESO  146-G14 with  a direct
method based 12+log(O/H)=7.61 \citep{Bergvall95} is a blue low surface
brightness galaxy  at a distance  of 21.4 Mpc.   DDO 68 with  a direct
method based  12+log(O/H)=7.21 \citep{Pustilnik05}  is a  blue compact
dwarf at a distance of 5.9 Mpc.  Ho II is a Magellanic irregular dwarf
at a distance of 3.3 Mpc  \citep{McCall12}, with a direct method based
12+log(O/H)=7.92 \citep{Croxall09}. 

 We also  compared our results  to the integrated measurements  of the
 KINGFISH sample  \citep{Kennicutt11}, the  DGS (Dwarf  Galaxy Survey)
 sample \citep{Madden13}, and other  two  EMP galaxies
 (I Zw 18 \& SBS 0335-052).  These properties are listed in Table~\ref{tab:properties}.

\begin{table}
\centering
\caption{\label{tab:properties} Properties of our four EMP galaxies along with additional two galaxies
  from the literature}
\begin{threeparttable}
\begin{tabular}{llll}
\hline
  Name     	  	&Distance   	&Metallicity          &Morphology\\
                		&[Mpc]          	&12+log(O/H)      &    \\
\hline
SextansA    	&1.4 			&7.49 	&dIrr\\
ESO146-G14 	&21.4   			&7.61 	&LSB   \\
DDO 68		&5.9 			&7.21 	&BCD\\
HoII       		&3.3				&7.92  	&dIrr\\
\hline
I Zw 18		&18.2\tnote{a} 	&7.14\tnote{b}  & BCD\\
SBS 0335-052 	&56.0\tnote{c} 	&7.25\tnote{d} & BCD\\
\hline
\end{tabular}
\begin{tablenotes}
        \item[a] \citet{Aloisi07}; 
        \item[b] \citet{Izotov99};
        \item[c] \citet{Madden13};
        \item[d] \citet{Izotov97}.     
      \end{tablenotes}
       \end{threeparttable}
\end{table}

\subsection{Observations}

 \begin{figure}
  \includegraphics[width=\columnwidth]{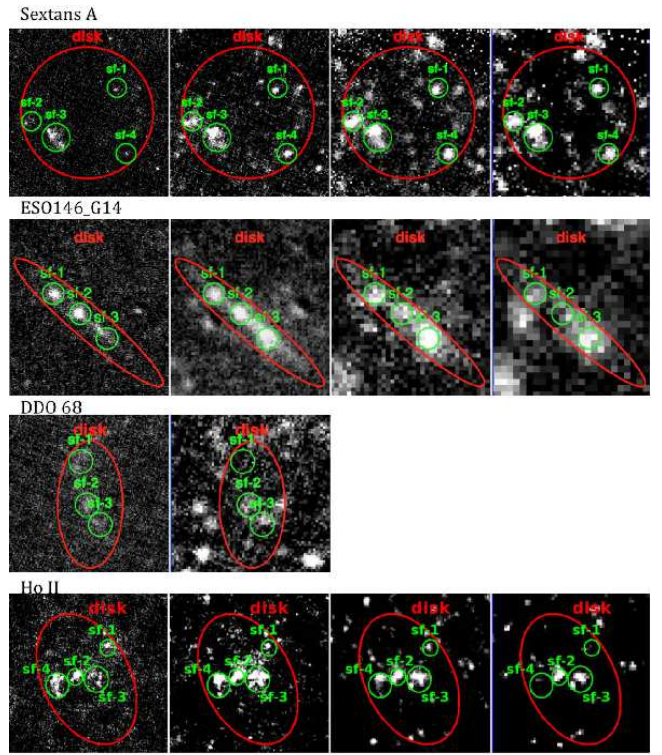}
  \caption{\label{mult_img} The Herschel images of our galaxies at 70, 160, 250 and 350 $\mu$m from left to right.
    The large ellipses indicate the extension of the whole star-forming disk and small circles are
    those individual dusty star-forming regions. }
\end{figure}

Herschel broad-band images (PI: Y. Shi, PID: OT2\_yshi\_3) of Sextans
A, ESO 146-G14 and DDO 68 were taken at 70 and 160 $\mu$m through the scan
map modes  of PACS \citep{Poglitsch10},  and 250, 350 and  500 $\mu$m
through the small  map modes with  SPIRE \citep{Griffin10}.  The exposure
time (excluding overheads) in the two PACS bands is 1.9 hr, 1.6 hr and
2.7 hr  for Sextans A,  ESO 146-G14 and  DDO 68, respectively,  and in the
three SPIRE bands is  6 min, 6 min and 10  min, respectively.  The data of
Sextans A and ESO 146-G14  have been partly published in \citet{Shi14}
where the detailed  data reduction was given, while the  data of DDO 68
is newly presented here. DDO 68 has low S/N, which may cause large uncertainties 
on the color (\S~\ref{sec_IR}). The PACS and SPIRE integration 
  times of the observations are also listed in in Table~\ref{tab:data}.
The Herschel data of Ho II was taken from the KINGFISH project \citep{Kennicutt11}. 

 In addition to the Herschel data, the Spitzer IR data at 24 $\mu$m
  and GALEX far-UV data  at 1516\AA were also retrieved from the archive. 

\begin{table}
\centering
\caption{\label{tab:data} Exposure time of {\it Herschel} data used in this paper.}
\begin{tabular}{ccc}
\hline
  Name     	  	&PACS   			&SPIRE \\
                		&70, 160 $\mu$m         &250, 350, 500 $\mu$m           \\
\hline
SextansA    	&1.9 hr 				&6 min \\
ESO146-G14 	&1.6 hr   				&6 min   \\
DDO 68		&2.7 hr 				&10 min \\
\hline
\end{tabular}
\end{table}

\subsection{Photometric Measurements}

\begin{table*}
  \scriptsize
  \centering
\caption{\label{tab:photometry} Photometry Of Individual Star-forming Regions In Metal Poor Galaxies}
\begin{threeparttable}
\begin{tabular}{llcccccccccccccccccc}
\hline
Region\tnote{a}     &   Position (J2000)\tnote{b}  &  m$_{a}$ $\times$m$_{b}$\tnote{c} &  f$_{\rm FUV}$\tnote{d}  & f$_{\it 3.6 \mu m}$    & f$_{\it 4.5\mu m}$ &  f$_{\it 24 \mu m}$ & f$_{\it 70 \mu m}$ & f$_{\it 100 \mu m}$  & f$_{\it 160 \mu m}$ & f$_{\it 250 \mu m}$ & f$_{\it 350 \mu m}$  \\ 
           &                              &  [arcsec$^{2}$]         &   [$\mu$Jy]     & [mJy]           & [mJy]        & [mJy]        & [mJy]        & [mJy]        & [mJy]        & [mJy]   \\ 
\hline
SextansA/sf-1            &10:10:56.9  -04:40:27.0 &22x22        &960  $\pm$0.4          &1.64$\pm$0.009           &1.05$\pm$0.008           &1.06$\pm$0.14            &   41$\pm$2     &      &   56$\pm$7           &   55$\pm$3           &   32$\pm$3           \\
SextansA/sf-2            &10:11:10.0  -04:41:44.5 &22x22        &660  $\pm$0.4          &1.65$\pm$0.009           &1.36$\pm$0.008           &3.22$\pm$0.33            &   72$\pm$3      &    &  147$\pm$18           &  111$\pm$4           &   52$\pm$4           \\
SextansA/sf-3            &10:11:6.20  -04:42:22.5  &32x32        &4300 $\pm$0.6          &2.78$\pm$0.013           &2.20$\pm$0.011           &6.30$\pm$0.64            &  267$\pm$4      &     &  297$\pm$24           &  164$\pm$5           &   89$\pm$4           \\
SextansA/sf-4            &10:10:55.5  -04:42:59.4 &22x22        &260  $\pm$0.4          &1.14$\pm$0.009           &0.69$\pm$0.008           &0.94$\pm$0.13            &   21$\pm$2       &    &   69$\pm$8           &   62$\pm$3           &   34$\pm$3           \\
ESO146-G14/sf-1          &22:13:6.0   -62:03:32.5  &10x10        &150  $\pm$0.2          &0.33$\pm$0.003           &0.24$\pm$0.003           &1.02$\pm$0.16            &   28$\pm$4     &      &   38$\pm$6           &   29$\pm$4           &   16$\pm$3           \\
ESO146-G14/sf-2          &22:13:2.5   -62:03:51.6  &10x10        &260  $\pm$0.2          &0.38$\pm$0.003           &0.28$\pm$0.003           &1.23$\pm$0.18            &   36$\pm$5      &     &   52$\pm$8           &   28$\pm$3           &   11$\pm$3           \\
ESO146-G14/sf-3          &22:12:59.0  -62:04:14.3 &10x10        &90   $\pm$0.2          &0.74$\pm$0.003           &0.50$\pm$0.003           &0.91$\pm$0.16            &   15$\pm$2      &     &   57$\pm$9           &   49$\pm$6           &   31$\pm$4           \\
DDO68/sf-a               &09:56:46.6  +28:50:16.6 &10x10        &340  $\pm$0.1          &0.08$\pm$0.004           &0.08$\pm$0.006           &0.71$\pm$0.08            &   11$\pm$1        &   &   $<$2                  &   $<$5                  &   $<$5                  \\
DDO68/sf-b               &09:56:46.2  +28:49:39.6 &10x10        &370  $\pm$0.1          &0.41$\pm$0.004           &0.25$\pm$0.006           &0.36$\pm$0.05            &   10$\pm$1        &   &    9$\pm$1           &   $<$5                  &   $<$5                  \\
DDO68/sf-c               &09:56:45.3  +28:49:22.6 &10x10        &350  $\pm$0.1          &0.60$\pm$0.004           &0.43$\pm$0.006           &0.33$\pm$0.05            &   10$\pm$1        &   &    8$\pm$1           &    7$\pm$1           &   $<$5                  \\
HoII/sf-1                &08:18:48.5  +70:44:40.1 &28x28        &880  $\pm$0.5          &1.34$\pm$0.008           &1.09$\pm$0.012           &16.69$\pm$1.67           &  319$\pm$6         &165$\pm$5 &  144$\pm$  14           &   79$\pm$4           &   51$\pm$4           \\
HoII/sf-2                &08:19:13.3  +70:42:56.3 &28x28        &2610 $\pm$0.5          &20.41$\pm$0.008          &14.19$\pm$0.012          &46.32$\pm$4.63           &  471$\pm$7       &497$\pm$5   &  407$\pm$  39           &  205$\pm$4           &  103$\pm$4           \\
HoII/sf-3                &08:18:57.2  +70:42:48.4 &46x46        &4740 $\pm$0.8          &12.31$\pm$0.014          &8.40$\pm$0.020           &17.91$\pm$1.80           &  449$\pm$7        &751$\pm$8   &  630$\pm$  19           &  345$\pm$7           &  196$\pm$6           \\
HoII/sf-4                &08:19:26.9  +70:42:27.0 &43x43        &4780 $\pm$0.7          &10.44$\pm$0.013          &7.99$\pm$0.018           &63.18$\pm$6.32           &  765$\pm$7        &892$\pm$7  &  387$\pm$  16           &  247$\pm$6           &  105$\pm$5           \\
\hline
\end{tabular}
\begin{tablenotes}
        \item[a] Star-forming regions we defined in our EMP galaxies;
        \item[b] RA \& DEC of the center of each region;
        \item[c] Major axis and minor axis of each region in arcsecond;
        \item[d] Flux density at far-UV of each region, the same with f$_{\it 3.6 \mu m}$, f$_{\it 4.5 \mu m}$, f$_{\it 70 \mu m}$, f$_{\it 160 \mu m}$, f$_{\it 250 \mu m}$, f$_{\it 350 \mu m}$.
      \end{tablenotes}
       \end{threeparttable}
\end{table*}

As detailed in \citet{Shi14}, the  star-forming disk of each galaxy is
defined as an ellipse/circle to  closely follow the 10$\sigma$ contour
of the far-UV emission, as shown in Fig.~\ref{mult_img}. Note that for
DDO 68 we  excluded the tidal tail  to better focus on  the main disk,
although  the   tail  is  formally  above   the  10$\sigma$  detection
threshold.  Individual dusty star-forming clumps within a star-forming
disk were  identified as circle  regions with elevated far-UV  and 160
$\mu$m IR  emission which is  3$\sigma$ above the fluctuations  of the
disk emission.  The  defined star-forming regions are  listed in Table
\ref{tab:photometry}.  For  the flux measurements, the  underlying sky
emission is  estimated within a sky  annuli between 1.1 and  1.5 times
the disk aperture.   For flux error estimates,  the {\it Herschel}
  flux uncertainty is given by the following sources: the first is the
  uncertainty of the photon noise  within the aperture we defined; the
  second is the  uncertainty from the sky  background subtraction; the
  third represents the  uncertainty introduced by the  offset from the
  accurate PSF position when we  defined the aperture for each source;
  and the system uncertainty is also included. For more details please
  see \citet{Shi14}.  The measurements of fluxes  at other wavelengths
  basically follow the same procedure.   The
  photometric results of  these star-forming regions are  listed in Table
  \ref{tab:photometry}.

\begin{table}
\centering
\scriptsize
\caption{\label{tab:pixsize} Pixel sizes of {\it Herschel} data used in this paper.}
\begin{threeparttable}
\begin{tabular}{cccccc}
\hline
Instrument			&Wavelength				&Pixel size              			&Resolution \\
\hline
{\it Herschel}/PACS 	&70, 160 $\mu$m 			&1\arcsec, 2\arcsec 			&5\arcsec, 13\arcsec \\
{\it Herschel}/SPIRE 	&250, 350, 500 $\mu$m		&4\arcsec, 6\arcsec, 8\arcsec	&18\arcsec, 25\arcsec , 36\arcsec\\
{\it Spitzer}/IRAC		&3.6, 4.5$\mu$m			&0.75\arcsec\tnote{a}, 0.75\arcsec&2.5\arcsec, 2.5\arcsec \\
{\it Spitzer}/MIPS		&24$\mu$m				&1.5\arcsec					&6\arcsec\\
{\it GALEX}/FUV		&1516$\mathring{\rm A}$					&1.5\arcsec					&4.5\arcsec\\
\hline
\end{tabular}
\begin{tablenotes}
        \item[a] The pixel size of the 3.6$\micron$ image of ESO146-G14 is 0.6$\arcsec$.       
      \end{tablenotes}
       \end{threeparttable}
\end{table}

 In order to compare images among different resolutions,
 aperture corrections are performed on all the star forming 
 clumps based on their PSFs. 
 We have tested that the aperture correction
  method has almost the same effect as convolving images to the same resolution
  when fitting SEDs \citep[e.g., see][]{Casasola15}. Pixel sizes and resolutions of images at each wavelength are listed in 
  Table~\ref{tab:pixsize}.	
				
 \section{The far-IR SEDs}\label{sec_IR}

\subsection{The color-color diagrams}

\begin{figure}
  \includegraphics[width=\columnwidth]{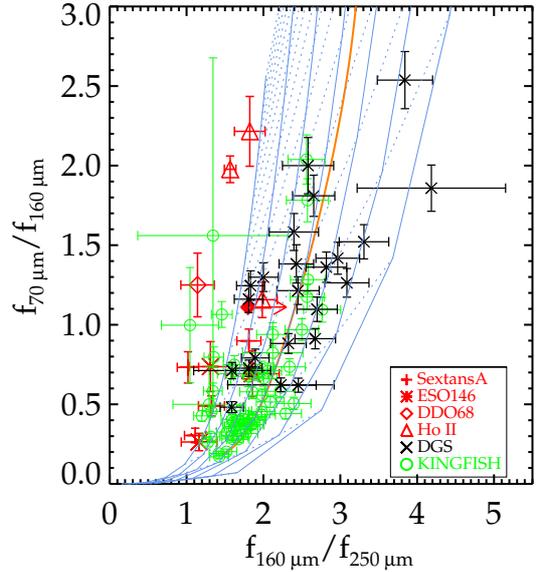}
\caption{\label{color_color_1}  The distribution  of EMP  star-forming
  regions,   and integrated measurements of galaxies from DGS and KINGFISH  in  the
  $f_{70{\mu}m}/f_{160{\mu}m}$ vs. $f_{160{\mu}m}/f_{250{\mu}m}$ color-color diagram.  The
  grid is a single modified black-body with a range of temperatures and
  emissivity indices. The solid line represents a constant emissivity index, $\beta$,
 with the temperature ranging from 10 to
100 K in  a decrement of 5 K from the  left-bottom to the upper-right. Dashed lines represent constant temperature
  with $\beta$ ranging from 0.0 to 3.0 in
a step of  0.5 from  the  left-bottom to  the upper-right.  The brown line is the trend of star forming galaxies predicted by 
 the model of \citet{Dale14}.}
\end{figure}

\begin{figure}
\includegraphics[width=\columnwidth]{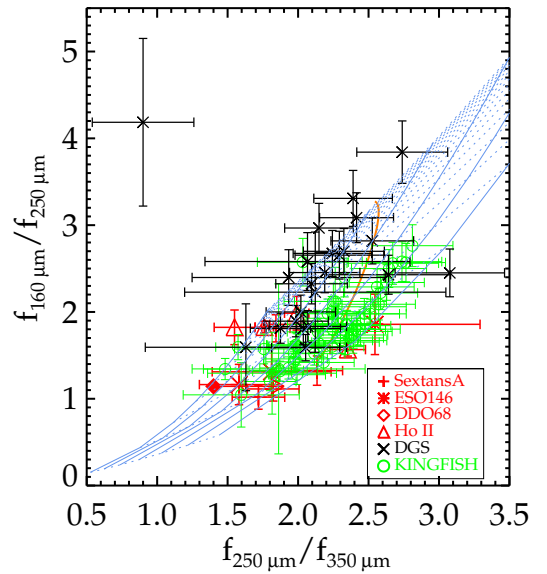}
\caption{\label{color_color_2} The same as Fig.\ref{color_color_1} but using
 $f_{160{\mu}m}/f_{250{\mu}m}$ vs. $f_{250{\mu}m}/f_{350{\mu}m}$. }
\end{figure}

Fig.~\ref{color_color_1}  shows  the  $f_{70{\mu}m}/f_{160{\mu}m}$  vs
$f_{160{\mu}m}/f_{250{\mu}m}$  color-color diagram  of our  individual
EMP regions along  with those integrated measurements  of the KINGFISH and
DGS galaxies.   The predictions  of single modified  black-body (MBB) models
are  overlaid as  grids where  the solid  line represents  a constant
emissivity index, $\beta$, with the temperature ranging from 10 to 100
K  in decrements of  5 K  from the  left-bottom to the upper-right, and the
dashed lines represent constant temperatures with $\beta$ ranging from 0.0 to 3.0
in steps of 0.5 from the left-bottom to the upper-right.  The included
DGS galaxies  are  those  with  Z/Z$_{\sun}$ above  $\sim$10\%,  because  those
 EMP ones in their  sample lack the enough  S/N ($\geq$ 3) to be
included.  As indicated by  Fig.~\ref{color_color_1}, our sample shows
a  quite  different behavior  as  compared  to  the DGS  and  KINGFISH
Spirals.  The majority  of the  KINGFISH  locates in  the region  with
$f_{70{\mu}m}/f_{160{\mu}m}$ $<$ 1.5 and $f_{160{\mu}m}/f_{250{\mu}m}$
$<$ 2.5, and  the DGS ranges from the bottom-left  to the upper-right,
confined  by the  overlaid  gray-body curves.   In  contrast, our  EMP
regions   mainly   occupy   the   left  side   of   the   plot,   with
$f_{160{\mu}m}/f_{250{\mu}m}$  $<$ 2  but $f_{70{\mu}m}/f_{160{\mu}m}$
covers a range from 0  to 2.5.  At given $f_{160{\mu}m}/f_{250{\mu}m}$
our     sample     shows     a      much     larger     scatter     in
$f_{70{\mu}m}/f_{160{\mu}m}$, and more  dramatically, a large fraction
of our  sample has $f_{70{\mu}m}/f_{160{\mu}m}$  above the limit  of a
single  modified black-body  with  $\beta$=0 (that  is the  black-body
emission)  for a  given $f_{160{\mu}m}/f_{250{\mu}m}$.   When compared
with the trend of star  forming galaxies predicted by empirical models
of \citet{Dale14}, the  median behavior of KINGFISH  and DGS basically
follow  the  trend, while  star-forming  regions  of our  EMP  galaxies
lie systematically above the  trend.  A reasonable explanation for
the  large   $f_{70{\mu}m}/f_{160{\mu}m}$  ratio   is  an   excess  of
$f_{70{\mu}m}$  emission  contributed  by   an  additional  source  of
heating.  The  stochastically heated  small grains may  be responsible
for the  excess 70 $\mu$m emission  \citep{Draine01}.  The requirement
of a  second heating  component when  performing a  fit to  the 70-500
$\mu$m  photometry has  been  widely  seen in  all  types of  galaxies
\citep{Galametz12,  Remy13}, while  our sample  seems to  be the 
extreme   case   in    which   about   half   of    our   sample   has
$f_{70{\mu}m}/f_{160{\mu}m}$ higher  than the  black-body curve  for a
given $f_{160{\mu}m}/f_{250{\mu}m}$ color.

Similar to Fig.~\ref{color_color_1}, Fig.~\ref{color_color_2} presents
another  color-color plot  that  is $f_{160{\mu}m}/f_{250{\mu}m}$  vs.
$f_{250{\mu}m}/f_{350{\mu}m}$.   Again,  our   EMP  regions  occupy  a
different  area of  the  diagram,  when compared  with  the other  two
sample.  The KINGFISH Spirals and DGS galaxies span a similar range of
$f_{250{\mu}m}/f_{350{\mu}m}$    but   the    latter   shows    larger
$f_{160{\mu}m}/f_{250{\mu}m}$  at given  $f_{250{\mu}m}/f_{350{\mu}m}$
compared to  the former.  In  contrast, our EMP regions  mainly occupy
the    locus   toward    the   bottom-left    corner   with    smaller
$f_{160{\mu}m}/f_{250{\mu}m}$ and  $f_{250{\mu}m}/f_{350{\mu}m}$. 
In  the  figure   there  is  one  object  (UM   461)  that  deviates
significantly  from the  trend. It  is difficult  to explain  such a
strange  SED. We  double  checked the  photometric measurements  and
obtained  consistent   values  with  those  in   the  literature  of
\citet{Remy13} as  used in the  figure. But  we did notice  that the
centroids of the 250 and 350 um images  are significantly offset
(14$''$) from  those of the  70 and 160 $\mu$m,  indicating possible
significant  contamination  by a  background  source  at these two
wavelengths.

\subsection{Modified black-body fitting}\label{sec-dust-MBB}

To further characterize the SED shape and understand the dust emission
of EMP  star-forming  regions, the MBB fittings are carried out.  A single
MBB model  fitting to the
full SED is not appropriate as  a single photon heating dust component
may contaminate  significantly the  70 $\mu$m or  even the  100 $\mu$m
emission. We  first performed a  single MBB fitting to  the photometry
$\geq$  100  $\mu$m  by  assuming  at  these  longer  wavelengths  the
contribution from the hot dust  component was negligible. The MBB model
determined  the dust  temperature  $T$ and  the  dust emissivity  index
$\beta$ from $S_{\nu} = A*B_{\nu}(T)\lambda^{-\beta}$, where $S_{\nu}$
was the  flux density  at frequency  $\nu$, $A$ was a constant  that is
related to  the column density  of the  dust, and $B_{\nu}(T)$  was the
Planck function.  This  method assumed an optically  thin condition at
the  observed  far-IR wavelengths  with  the  dust mass  derived  from
$M_{\rm dust} = \frac{S_{\nu} \rm D^{2}}{ \rm \kappa_{\nu} B_{\nu}(\it
  T)}$,  where $\kappa_{\nu}$  was the  dust  opacity, and  $D$ was  the
distance  to  the  galaxy.   We took  $\kappa_{\nu}$  =  1.9  cm$^{2}$
g$^{-1}$                 at                $\lambda$                 =
350$\mu$m \footnote{http://www.astro.princeton.edu/$\sim$draine/dust/dustmix.html,
  (Milky    Way,   R\_V    =    3.1),    \citet{Li01}}, and calculated the
dust mass at 
$\lambda$=350$\mu$m  as this band was less sensitive
to temperature  than shorter wavelengths, and  had lower uncertainties
than   those   at   longer    wavelengths.    As   listed   in   Table
\ref{tab:one_mbb}, the  single MBB fitting  gave a $\beta \sim  $ 1
and  $T$  $\sim$ 20-50 K.   The  temperature  may be  underestimated  for
Sextans A and ESO 146-G14 as there is no 100 $\mu$m photometry.

\begin{figure}
\includegraphics[width=\columnwidth]{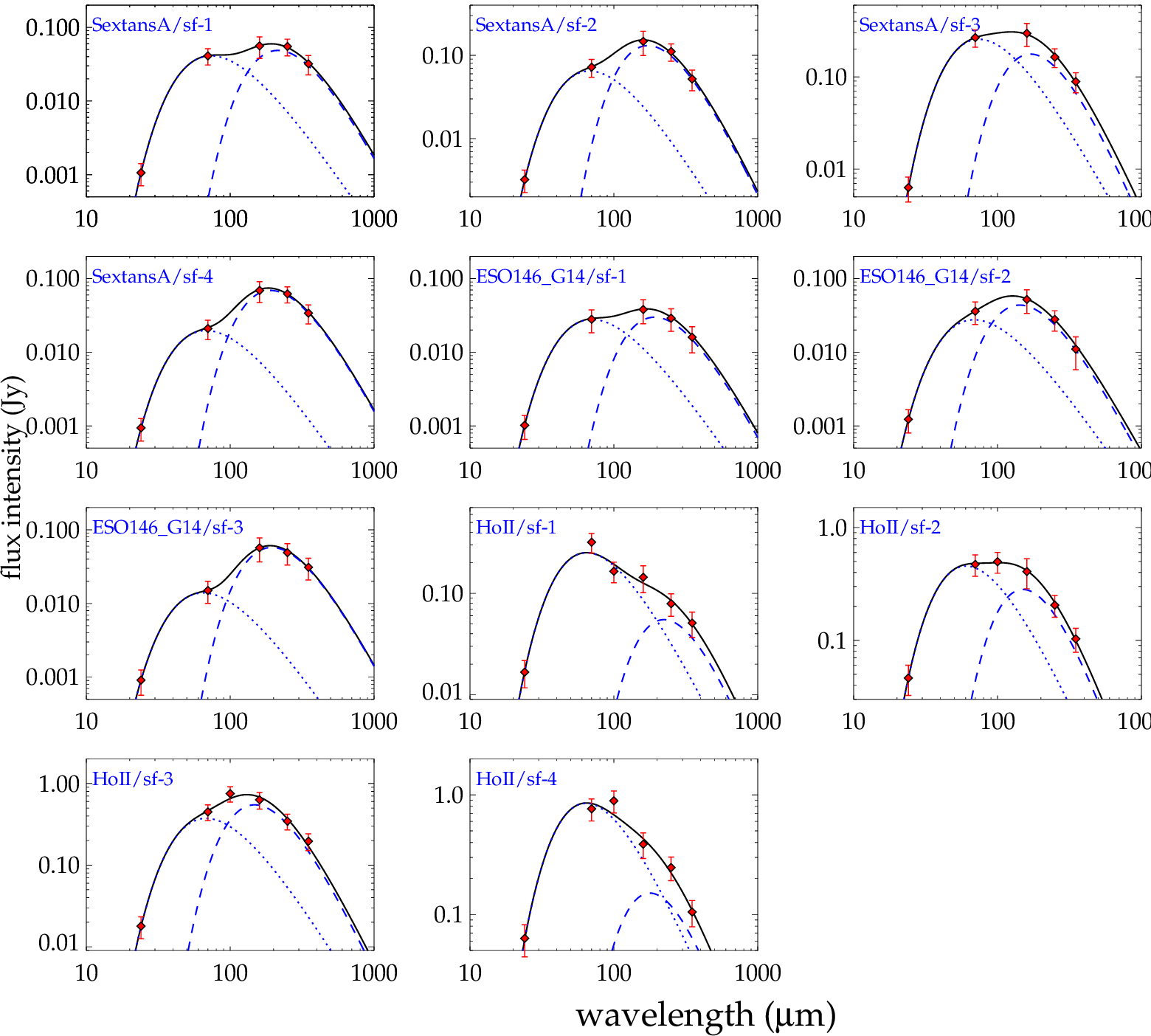}
\caption{\label{fig:two_mbb} The two modified black-body fitting to the SEDs. 
Reds dots are the flux density at different wavelength within each star-forming 
region. Black solid lines are  the  2T MMB best fits with blue dotted lines the 
best fits of the warm dust components and blue dashed lines the best fits of 
the cold dust components. }
\end{figure}

The two-MBB fitting is more reasonable  by taking the advantage of the
full  SED,  which  is  carried   out  with  the  equation  $S_{\nu}  =
A_{w}*B_{\nu}(T_{w})\lambda^{-\beta_{w}}                             +
A_{c}*B_{\nu}(T_{c})\lambda^{-\beta_{c}}$, where  $A_{\rm w}$, $T_{\rm
  w}$  and $\beta_{\rm  w}$ describe  the  hot component  of the  dust
emission while  $A_{\rm c}$, $T_{\rm  c}$ and $\beta_{\rm c}$  are for
the cold component.  Here we fixed $\beta_{w}$ = 1  to represent the hot
small grains and $\beta_{c}$ = 2 for the cold large grains, similar to
the   study   of   \citet{Zhu09}.     The   results   are   shown   in
Fig.~\ref{fig:two_mbb} and listed in Table~\ref{tab:two_mbb}. The dust
mass is derived in the same way as the single MBB fitting. If the cold
component dust  indices $\beta_{c}$ decrease  from 2.0 to 1.7,  1.5 and
1.3, the derived $T_{\rm  c}$ increases and the derived dust mass drops 
by 10\%, 25\% and 50\%, respectively.

\begin{figure}
\includegraphics[width=\columnwidth]{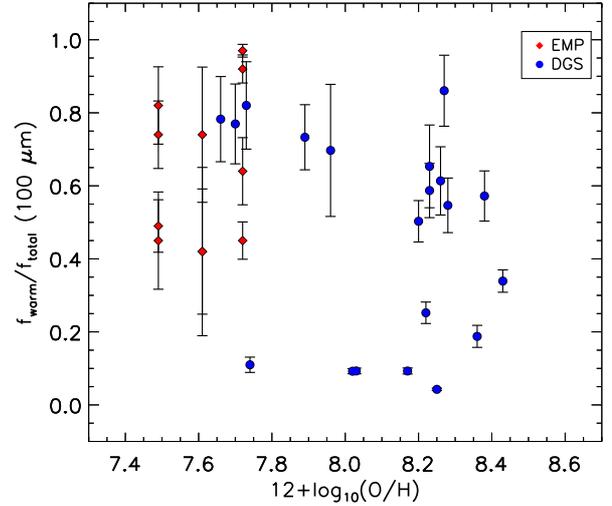}
\caption{\label{fracwarm_Z} The fraction of warm dust emission at 100 $\mu$m  as a function of the
  Oxygen abundance based on the two modified black-body fitting.  The red diamonds are the
  EMP star-forming regions, and blue circles are for the integrated dwarf galaxies from the DGS. }
\end{figure}

Our two MBB fits give cold dust with $T_{\rm c} \sim$ 15-20 K and warm
dust with $T_{\rm w}  \sim$ 50-60 K.  Fig.~\ref{fig:two_mbb} indicates
that the warm dust emission  is significant for these EMP star-forming
regions,  in  contrast to  those  of  Spirals \citep{Galametz12}.   To
further  characterize the  importance  of the  hot  dust emission,  we
performed  similar fittings  to the  integrated SEDs  of the  DGS sample
\citep{Remy13} and investigated the fraction  of the emission from the
warm    dust    MBB    at   100    $\mu$m    $(f_{\rm    warm}/f{_{\rm
    total}})_{100{\mu}m}$ as a function  of the oxygen abundance.  The
result  is shown  in Figure~\ref{fracwarm_Z}.   It indicates  that the
fraction  $(f_{\rm  warm}/f{_{\rm total}})_{100{\mu}m}$  increases  on
average with decreasing metallicities: around the Solar abundance, the
warm dust  emission is small ($<$  20\%) but reaches above  50\% below
one  tenth  of   the  Solar  metallicity.   Such  a   high  warm  dust
contribution cautions the single MBB fitting to the photometry for EMP
galaxies when the data at  $\lambda$ $\le$ 100 $\mu$m are included.  As
shown in Figure~\ref{fig:two_mbb}, our two-MBB fittings with $\beta$=2
for the cold component also provide good fittings, while the single
MBB requires  a lower  $\beta$.  This suggests  that the  result about
$\beta$ is sensitive to the way  how the MBB fitting is performed, and
the $\beta$ of  metal poor galaxies can be underestimated  as a result
of a large warm dust contribution up to 100 $\mu$m.

\begin{table}
\centering
\scriptsize
\caption{\label{tab:one_mbb} Results of Single MBB Fits to the SEDs at $\geq$ 100 $\mu$m}
\begin{tabular}{lllllllllll}
\hline
  Region     &  T    & $\beta$  & M$_{\rm dust}$ \\
             &  [K]  &          & [M$_{\odot}$]\\
\hline
SextansA/sf-1     &   19$\pm$  8   &   0.99$\pm$1.22  &   1.34$\times10^{3}$\\
SextansA/sf-2     &   25$\pm$  9   &   1.02$\pm$0.78  &   1.31$\times10^{3}$\\
SextansA/sf-3     &   37$\pm$ 17   &   0.59$\pm$0.56  &   9.93$\times10^{2}$\\
SextansA/sf-4     &   21$\pm$  9   &   1.00$\pm$1.12  &   1.21$\times10^{3}$\\
ESO146-G14/sf-1   &   23$\pm$ 22   &   0.90$\pm$2.24  &   1.00$\times10^{5}$\\
ESO146-G14/sf-2   &   30$\pm$  5   &   1.14$\pm$0.64  &   4.51$\times10^{4}$\\
ESO146-G14/sf-3   &   22$\pm$ 16   &   0.80$\pm$1.94  &   2.11$\times10^{5}$\\
HoII/sf-1         &   47$\pm$ 13   &   0.07$\pm$0.44  &   1.98$\times10^{3}$\\
HoII/sf-2         &   37$\pm$  4   &   0.70$\pm$0.24  &   6.03$\times10^{3}$\\
HoII/sf-3         &   33$\pm$  1   &   0.82$\pm$0.06  &   1.30$\times10^{4}$\\
HoII/sf-4         &   41$\pm$  1   &   0.95$\pm$0.06  &   5.42$\times10^{3}$\\
\hline
\end{tabular}
\end{table}

\begin{table*}
\scriptsize
\caption{\label{tab:two_mbb} Results of two MBB Fits to the Photometry}
\begin{tabular}{lllllllll}
\hline
  Region     &  T$_{\rm cold}$    & T$_{\rm warm}$   & M$_{\rm warm}$ & M$_{\rm cold}$ & $M_{\rm warm}$/$M_{\rm cold}$  & $f_{\rm warm}$/$f_{total}$(100$\mu$m) & $L_{8-1000{\mu}m}$ \\
             &  [K]            &   [K]        & [M$_{\odot}$]      & [M$_{\odot}$]     &                                    &                & [L$_{\odot}$] \\
\hline
SextansA/sf-1         & 13$\pm$2  &   49$\pm$2  &2.22$\times10^{1}$&   3.33$\times10^{3}$& 6.7$\times10^{-3}$&  0.82    &      1.93$\times10^{5}$\\
SextansA/sf-2         & 16$\pm$2  &   54$\pm$3  &2.49$\times10^{1}$&   3.00$\times10^{3}$& 8.3$\times10^{-3}$&  0.49    &      4.09$\times10^{5}$\\
SextansA/sf-3         & 17$\pm$3  &   49$\pm$2  &1.43$\times10^{2}$&   3.72$\times10^{3}$& 3.8$\times10^{-2}$&  0.74    &      1.11$\times10^{6}$\\
SextansA/sf-4         & 15$\pm$1  &   53$\pm$3  &7.74$\times10^{0}$&   2.62$\times10^{3}$& 3.0$\times10^{-3}$&  0.45    &      1.53$\times10^{5}$\\
ESO146-G14/sf-1       & 15$\pm$2  &   52$\pm$2  &2.78$\times10^{3}$&   2.60$\times10^{5}$& 1.0$\times10^{-2}$&  0.74    &      2.98$\times10^{7}$\\
ESO146-G14/sf-2       & 20$\pm$4  &   53$\pm$5  &2.50$\times10^{3}$&   8.64$\times10^{4}$& 2.9$\times10^{-2}$&  0.42    &      3.77$\times10^{7}$\\
ESO146-G14/sf-3       & 14$\pm$2  &   56$\pm$3  &1.00$\times10^{3}$&   5.35$\times10^{5}$& 1.9$\times10^{-3}$&  0.42    &      2.66$\times10^{7}$\\
HoII/sf-1             & 13$\pm$3  &   56$\pm$2  &4.19$\times10^{2}$&   2.39$\times10^{4}$& 1.8$\times10^{-2}$&  0.97    &      5.46$\times10^{6}$\\
HoII/sf-2             & 19$\pm$3  &   60$\pm$4  &5.81$\times10^{2}$&   1.69$\times10^{4}$& 3.4$\times10^{-3}$&  0.64    &      1.21$\times10^{7}$\\
HoII/sf-3             & 19$\pm$2  &   53$\pm$3  &7.68$\times10^{2}$&   3.04$\times10^{4}$& 2.5$\times10^{-2}$&  0.45    &      1.17$\times10^{7}$\\
HoII/sf-4             & 16$\pm$4  &   57$\pm$2  &1.36$\times10^{3}$&   2.11$\times10^{4}$& 6.4$\times10^{-2}$&  0.92    &      1.88$\times10^{7}$\\
\hline
\end{tabular}
\end{table*}

\subsection{Spatial variations of  SEDs And Dust Heating Mechanism}

\begin{figure*}
\includegraphics[width=2\columnwidth]{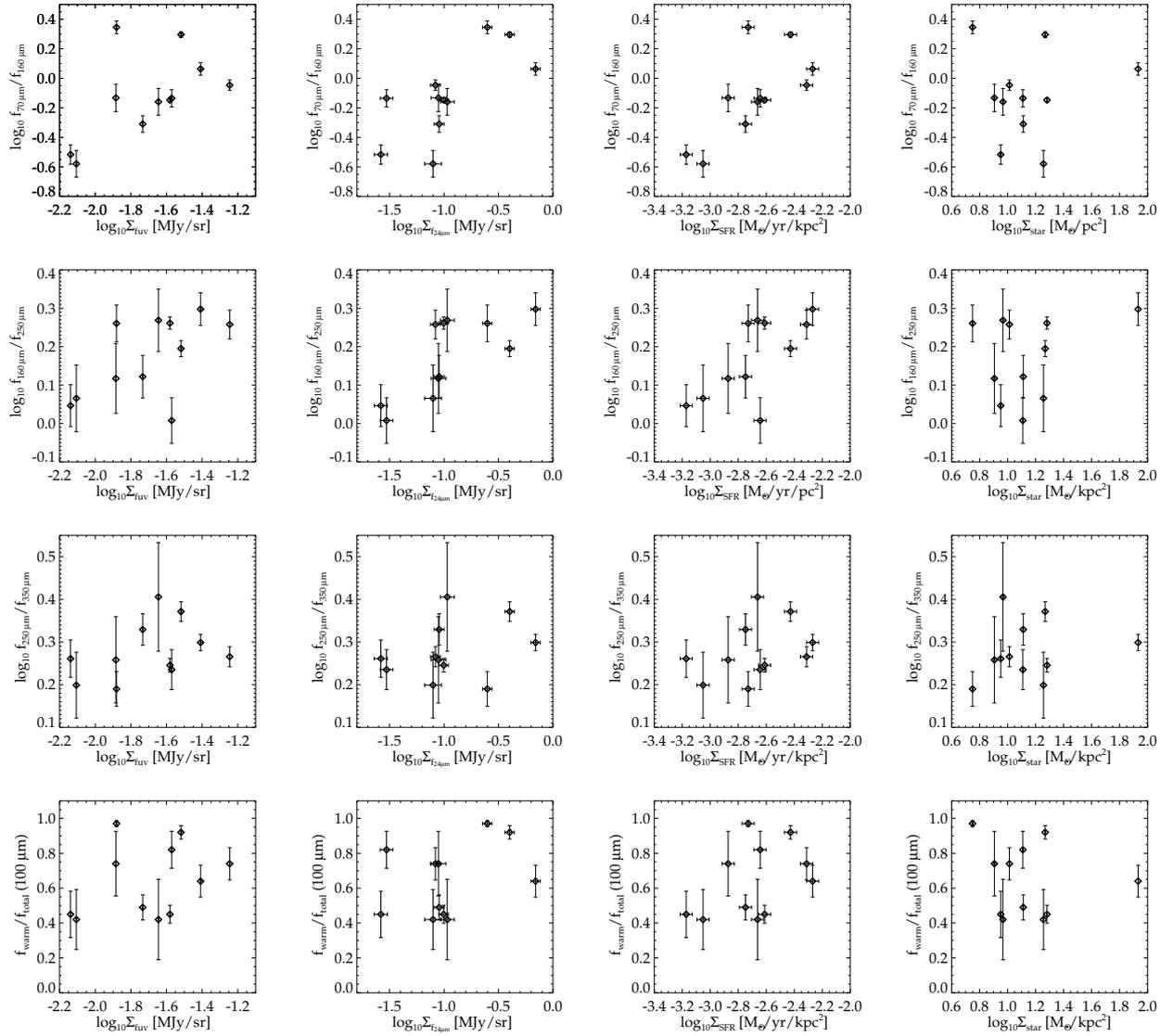}
\caption{\label{fig:trend_SB} The IR colors as functions far-UV surface brightness,
24 $\mu$m surface brightness, SFR surface densities and stellar mass surface densities from left to right. }
\end{figure*}

Spatially  resolved IR  data  allow to  investigate the  spatial
variation of the IR SEDs among  different star-forming regions in our 
EMP galaxy sample.  As  shown in  Fig.~\ref{color_color_1}, the variation  in the
color $f_{70{\mu}m}/f_{160{\mu}m}$  among different EMP regions  for a
given galaxy seems to be smaller as compared to the overall scatter of
the  integrated  color.  The
$f_{70{\mu}m}/f_{160{\mu}m}$ spans an overall  range from about 0.2 to
2.5.    In   contrast,   four   EMP  regions   in   Sextans   A   have
$f_{70{\mu}m}/f_{160{\mu}m}$ between  0.5 and 1.0, four  regions of Ho
II span  the color  range from 0.7  to 2.3, and  three regions  of ESO
146-G14   show   the  color   from   0.2   to   0.7.   As   shown   in
Fig.~\ref{color_color_2},     the    variation     in    the     color
$f_{160{\mu}m}/f_{250{\mu}m}$  among different  EMP  regions within  a
given galaxy is also smaller than the overall scatter among integrated
quantities of galaxies.

With the spatially resolved images, we  can further study the color as
a   function   of   local   conditions    in   the   2-D.    In   Fig.
\ref{fig:trend_SB},  the far-IR  color  of individual  EMP regions  is
investigated  as   a functions  of  far-UV surface brightness,
   24  $\mu$m   surface   brightness,   SFRs  and   stellar
masses\footnote{  SFRs  and  stellar masses  are  estimated  following
  \citet{Leroy08}'s  formula.  SFR from \citet{Leroy08} is calibrated from 
  Spiral galaxies and combines both UV (unobsured) and IR (obscured) 
 maps, which is applicable to low-metallicity galaxies \citep{Calzetti07, Salim07}. 
 For one galaxy (Sextans A), we collected the broad-band photometry 
 (FUV, NUV, V-band, 3.6 $\mu$m and 4.5 $\mu$m) and derived stellar masses
  based on the SED fitting. The result is not that different from the one 
  based on  \citet{Leroy08}'s method, with the difference < 20\%.}.
      All  surface brightness/densities  as
listed in  Table.~\ref{tab:SB} are inclination corrected  with angles of
0$^{\circ}$, 50$^{\circ}$,  54$^{\circ}$ and 30$^{\circ}$  for Sextans
A, ESO 146-G14,  DDO 68 and Ho II, respectively.   As indicated by the
figure, the far-UV  surface brightness of our EMP  regions are between
about  0.01 and  0.1  MJy/sr, which  is within  the  range of  Spirals
\citep{Gil-de-Paz07,  Shi11}.  The  SFR surface  densities of  our EMP
regions      are      between      $10^{-3.5}$      and      10$^{-2}$
M$_{\odot}$/yr/kpc$^{2}$.    This  is   also  within   the  range   of
star-forming regions  of local Spirals \citep{Bigiel08}.   The stellar
mass  surface  densities   of  EMP  regions  are  between   5  and  15
M$_{\odot}$/kpc$^{2}$,  which  is several  times  lower  than those  of
Spirals \citep[e.g.][]{Shi11}.

Fig.~\ref{fig:trend_SB} shows  that all  three IR  colors including
$f_{70{\mu}m}/f_{160{\mu}m}$,     $f_{160{\mu}m}/f_{250{\mu}m}$    and
$f_{250{\mu}m}/f_{350{\mu}m}$ increase on  average with the increasing
far-UV surface brightness, 24$\mu$m surface brightness and SFR surface
densities, with the best relationships with the SFR surface densities.
While all these three surface brightness/densities are associated with
young  stars,  the  SFR  derived here  includes  both  the  unobscured
component as  traced by the far-UV  and the obscured one  as traced by
the 24 $\mu$m.  
Fig.   \ref{fig:trend_SB}
further shows that there are no trends between the IR colors and
the stellar mass surface densities.   Unlike the SFR, the stellar mass
is better related  to old stars.  Investigations  of the relationships
between the far-IR colors and the  tracers 
can be used to constrain if the dust is mainly heated by the radiation
from the young stars or the  interstellar radiation field from the old
stars.  As  argued by  \citet{Bendo15}, such investigations  show many
advantages in understanding the dust  heating mechanism as compared to
other methods such as the dust  SED fitting and radiative transfer that
relies on  assumptions of dust grain  properties, SED  shapes of
heating  source  etc.  The  results  of  Fig. \ref{fig:trend_SB}  thus
support that young stars are the main heating source of dust radiating
at wavelengths from 70 $\mu$m to $\geq$ 250 $\mu$m in EMP star-forming
regions. Investigations  of the heating  source of dust in  Spirals by
\citet{Bendo15}  found  that  in  only  3  out  of  24  galaxies,  the
$f_{160{\mu}m}/f_{250{\mu}m}$  and  $f_{250{\mu}m}/f_{350{\mu}m}$  are
better related  to the SFRs than  old stars while in  the remaining the
$f_{250{\mu}m}/f_{350{\mu}m}$ color  is driven by both.

The above result suggests that the dust of emission at 70 $\mu$m up to
 350 $\mu$m in EMP regions is heated by young stars instead of
diffuse stellar  radiation from old  stars.
The  underlying  cause  for  this needs  further  investigations.   As
discussed above, the SFR surface densities  of our EMP regions are not
that different from those in  Spirals, indicating the radiation fields
from  young  stars  are  not  enhanced  in  EMP  regions  compared  to
star-forming regions in  Spirals. But the SFR relative  to the stellar
mass, i.e. the specific SFR (sSFR),  in these EMP regions is enhanced,
with the  median of log(sSFR[Gyr$^{-1}$])  around -0.82 and  a standard
deviation  of  0.36. In  contrast,  the  star-forming regions  of  12
Spirals in \citet{Shi11} have the median log(sSFR[Gyr$^{-1}$]) of -1.28
and a standard  deviation of 0.32.   Compared to the sSFRs of dwarf 
galaxies as studied by \citet{Hunt15}, our galaxies lie below their trend,
 but still within the scatter  (0.01 -- 30 Gyr$^{-1}$, 
 around metallicities of our galaxies). This difference could 
 be due to the fact that our galaxies are mostly dwarf irregulars 
 and low-surface-brightness galaxies, while the sample of \citet{Hunt15} 
 contains many blue compact dwarf galaxies that are known to 
 be compact with enhanced SFRs and have higher sSFRs due to  interactions. 

The properties of dust grains  in EMP galaxies may be  systematically different from
those in Spirals, making the heating from young stars important to the
dust emission  all the way  up to 350  $\mu$m.  For example,  if small
dust grains  in EMP  galaxies are  abundant, they  could be  heated to
higher temperatures without requiring an enhanced radiation field.  It
is   observationally  difficult   to   quantify  the   size  of   dust
grains. Studies of extinction curves  point out that metal poor dwarfs
in the local  group, SMC and LMC, show more  steeply rising extinction
at  UV  wavelengths,  suggesting  smaller dust  grains  in  these  two
galaxies \citep[for a review, see][]{Li15}.  A simple extrapolation of
this result into  the EMP regime  would naturally expect the dust in  EMP galaxies to
be smaller  due to the lack  of raw material for  grain growth.  Small
grains could  offer more surface  area for the formation  of molecular
hydrogen  that is  likely abundant  in  EMP galaxies  as indicated  by
several  tracers including  dust  \citep[e.g.][]{Shi14}, warm  H$_{2}$
\citep[e.g.][]{Hunt10} and  [CII] 158  $\mu$m \citep[e.g.][]{Madden97,
  Madden13} although CO is  very weak \citep[e.g.][]{Shi15}.  However,
it is impossible with current facilities to obtain the measurements of
the extinction  curve for these  relatively distant EMP  galaxies, and
thus to conclude the size of dust grains in EMP galaxies.

\section{Dust-to-stellar mass ratio}\label{sec_mass}

\begin{figure}
\includegraphics[width=\columnwidth]{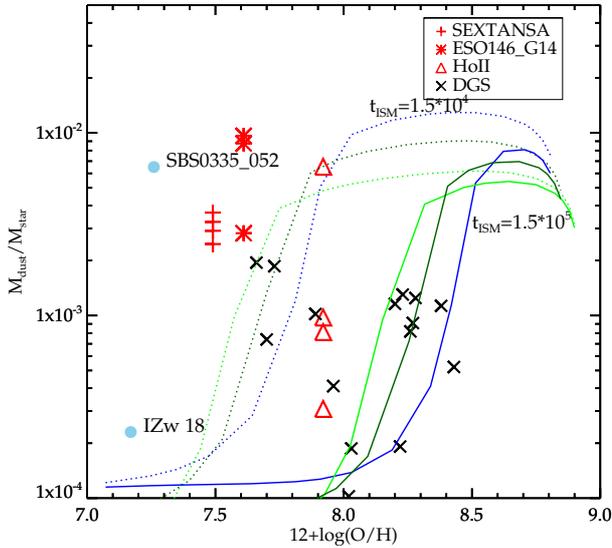}
\caption{\label{Mdust2Mstar_Z} The dust-to-stellar mass ratio of our EMP star-forming
  regions and integrated galaxies of the DGS. 
  The lines are the model predictions by \citet{Feldmann15}. 
  $t_{ISM}$ is the dust growth time-scale which  can be computed from basic collision theory \citep{Weingartner99} (solid or dashed lines). 
  $Q_{MS}$ denotes the (multiplicative) offset of a given galaxy from the main sequence (blue: 1/3, cyan:  1, green: 3).
  The data for IZw 18 and SBS 0335-052 are from \citet{Hunt14}.  }
\end{figure}

In Fig.~\ref{Mdust2Mstar_Z}, the dust-to-stellar mass ratio of our EMP
regions are  plotted against the oxygen  abundance, 12+log(O/H), along
with the integrated quantities from the  DGS.  The dust masses of both
samples are measured using the same  method, i.e.  the two MBB fitting
(see \S\ref{sec-dust-MBB}),  which could avoid the  artificial effects
of  different dust  mass estimates  on the  relationship.  As shown in  the figure, there
does not seem  to exist any correlation between  these two quantities.
\citet{Hunt14}  also found  that  two  EMP galaxies  (IZw  18 and  SBS
0335-052) show very different dust-to-stellar mass ratio although both
are similarly very metal poor, and their ratios lie within the overall
scatter  of spiral  and  other dwarfs  at  higher metallicities.   Our
significantly increased number of the data-points in the EMP regime as
compared to the work of  \citet{Hunt14} allows us to derive conclusive
results  that the  dust-to-stellar mass  ratio is  not related  to the
metallicity. The  figure further  indicates that  even within  a given
galaxy, the dust-to-stellar mass ratio of individual EMP regions could
show large scatters,  e.g.  EMP regions in Ho II  span the whole range
from roughly 10$^{-4}$ to 10$^{-2}$ while EMP regions in Sextans A and
ESO 146-G14 have much smaller scatters.

The overlaid lines in Fig.~\ref{Mdust2Mstar_Z} are the predictions of
the  model by  \citet[private  communication]{Feldmann15}, where t$_{\rm ISM}$
 is the dust growth time-scale which  can be derived from \citet{Weingartner99}'s 
 basic collision theory  (solid or dashed lines).
 Blue, cyan and green lines refer to Q$_{\rm MS}$ = $\frac{1}{3}$, 1,  3,
  which denote the (multiplicative) offset of a given galaxy from the main sequence.
  While  the normalization of  the trend depends  on the methodology used  for dust
mass and  metallicity measurements,  the model-predicted  sharp drop   
between  1/10 and  1/5 solar  abundance is  clearly not  seen in
Fig.~\ref{Mdust2Mstar_Z}.  The model of \citet{Feldmann15} invokes gas
outflow, inflow and star formation in an equilibrium state in order to
reproduce  the  observed  dust-to-gas  mass ratio  as  a  function  of
metallicity  as well  as   many other observed  galaxy properties.  The
model  is  motivated  to  explain  the  observed  sharp  drop  in  the
dust-to-gas  mass  ratio  vs.    the  metallicity  around  20\%  solar
metallicity \citep{Remy14,  Shi14}. Fig.~\ref{Mdust2Mstar_Z} indicates
that although a sharp drop may occur in the dust-to-gas  ratio, a 
similar drop in the trend of the dust-to-stellar ratio as a function of metallicity is not seen. The
model argues that the dust content in EMP galaxies is mainly regulated
by the galactic outflow, whose efficiency may be overestimated so that
too much dust is removed relative to the stellar content.
Our results further suggest that if outflows regulate the dust-to-stellar 
mass ratio, they must vary greatly on scales of 100 -- 1000 pc.

\begin{table*}
\centering
\scriptsize
\caption{\label{tab:SB} Measurements of 2-D Densities of EMP Star-Forming Regions }
\begin{tabular}{lccccccccccccccccccc}

\hline
  Region     &  $\Sigma_{\rm fuv}$  	& $\Sigma_{\rm 24\mu m}$ 	& $\Sigma_{\rm SFR}$   	& $\Sigma_{\rm star}$ 	& $\Sigma_{\rm dust}$  \\
                  &  [MJy/sr]  			&    [MJy/sr]        			& $\rm [M_{\odot}/yr/kpc^{2}]$ 	& $\rm [M_{\odot}/pc^{2}]$	& $\rm [M_{\odot}/kpc^{2}]$\\
\hline
SextansA/sf-1     &        (2.69$\pm$0.018)$\times10^{-2}$  &      (2.97$\pm$0.39)$\times10^{-2}$    &     (2.27$\pm$0.23)$\times10^{-3}$    &    (1.28$\pm$0.007)$\times10^{1}$  &      (4.72$\pm$ 0.47)$\times10^{4}$\\
SextansA/sf-2     &        (1.85$\pm$0.026)$\times10^{-2}$  &      (9.01$\pm$0.92)$\times10^{-2}$    &     (1.78$\pm$0.18)$\times10^{-2}$    &   (1.29$\pm$0.007)$\times10^{1}$  &      (3.75$\pm$ 0.38)$\times10^{4}$\\
SextansA/sf-3     &        (5.69$\pm$0.061)$\times10^{-2}$   &     (8.33$\pm$0.85)$\times10^{-2}$   &      (4.87$\pm$0.49)$\times10^{-3}$   &     (1.03$\pm$0.005)$\times10^{1}$  &      (3.85$\pm$ 0.39)$\times10^{4}$\\
SextansA/sf-4     &        (7.27$\pm$0.668)$\times10^{-3}$   &     (2.63$\pm$0.36)$\times10^{-2}$    &     (6.73$\pm$0.67)$\times10^{-4}$    &    (8.93$\pm$0.070)$\times10^{0}$   &     (3.79$\pm$0.38)$\times10^{4}$\\
ESO146-G14/sf-1   &    (1.31$\pm$0.058)$\times10^{-2}$  &      (8.88$\pm$0.22)$\times10^{-2}$     &    (1.34$\pm$0.13)$\times10^{-3}$  &      (8.04$\pm$0.073)$\times10^{0}$  &      (9.84$\pm$0.98)$\times10^{4}$\\
ESO146-G14/sf-2  &     (2.26$\pm$0.033)$\times10^{-2}$  &      (1.07$\pm$0.24)$\times10^{-1}$   &      (2.18$\pm$0.22)$\times10^{-3}$   &     (9.26$\pm$0.073)$\times10^{0}$    &    (3.59$\pm$0.36)$\times10^{4}$\\
ESO146-G14/sf-3   &    (7.83$\pm$0.096)$\times10^{-3}$   &     (7.92$\pm$0.22)$\times10^{-2}$   &      (8.88$\pm$0.89)$\times10^{-4}$   &    (1.80$\pm$0.007)$\times10^{1}$   &      (1.80$\pm$0.18)$\times10^{5}$\\
HoII/sf-1         &             (1.32$\pm$0.025)$\times10^{-2}$    &    (2.50$\pm$0.29)$\times10^{-1}$   &      (1.87$\pm$0.29)$\times10^{-3}$  &      (5.61$\pm$0.034)$\times10^{0}$    &    (6.50$\pm$0.65)$\times10^{5}$\\
HoII/sf-2        &              (3.90$\pm$0.008)$\times10^{-2}$   &     (6.93$\pm$0.80)$\times10^{-1}$     &    (5.38$\pm$0.54)$\times10^{-3}$  &      (8.55$\pm$0.003)$\times10^{1}$  &      (4.01$\pm$0.40)$\times10^{4}$\\
HoII/sf-3         &             (2.63$\pm$0.007)$\times10^{-2}$     &   (9.93$\pm$0.12)$\times10^{-2}$&         (2.45$\pm$0.25)$\times10^{-3}$  &      (1.91$\pm$0.002)$\times10^{1}$ &       (3.38$\pm$0.34)$\times10^{4}$\\
HoII/sf-4         &             (3.03$\pm$0.006)$\times10^{-2}$   &     (4.01$\pm$0.46)$\times10^{-1}$  &       (3.74$\pm$0.37)$\times10^{-3}$   &     (1.85$\pm$0.002)$\times10^{1}$ &       (2.76$\pm$0.28)$\times10^{4}$\\
\hline
\end{tabular}
\end{table*}


\section{Conclusions}\label{sec_conclusion}

  We present IR SEDs of  individual star-forming regions in four
  EMP galaxies observed by {\it Herschel}.  The main conclusions are:
  
  (1) As  compared to  spirals and  higher metallicity
  dwarfs,  EMP  star-forming regions have on  average much higher
  $f_{70{\mu}m}/f_{160{\mu}m}$         ratios         at         given
  $f_{160{\mu}m}/f_{250{\mu}m}$  ratios.  In  addition,  single  MBB
  fits to  the SED at  $\lambda \ge$  100 $\mu$m show  higher dust
  temperatures and lower emissivity indices, while two-MBB fits 
   with a  fixed emissivity index  show that even  at 100
  $\mu$m about half of the  emission comes from warm ($\sim$ 50 K)
  dust,  unlike that seen in Solar metallicity spiral galaxies.
    
(2)  Our spatially  resolved  images further  reveal  that the  far-IR
  colors            including            $f_{70{\mu}m}/f_{160{\mu}m}$,
  $f_{160{\mu}m}/f_{250{\mu}m}$ and  $f_{250{\mu}m}/f_{350{\mu}m}$ are
  all related  to the surface  densities of young stars  (far-UV, 24
  $\mu$m  and   SFRs),  but   not  with   the  stellar   mass  surface
  densities. This suggests that the  dust emitting at wavelengths from
  70 $\mu$m all  the way up to  the 350 $\mu$m is  heated by radiation
  from young stars instead of old stars.
  
  (3) Our EMP regions cover a  large range in the dust-to-stellar mass
  ratio, indicating  the importance of local  conditions, such as outflows 
   etc., in regulating the dust content.

\section*{Acknowledgements}

 We thank the anonymous referee  for helpful suggestions that improved
 the quality of the paper.  L.Z. and Y.S. acknowledge support for this
 work from  the National  Natural Science  Foundation of  China (grant
 11373021), the Strategic Priority  Research Program "The Emergence of
 Cosmological Structures"  of the  Chinese Academy of  Sciences (grant
 No.  XDB09000000),   and  Excellent   Youth  Foundation   of  Jiangsu
 Scientific Committee  (grant BK20150014) .  L.Z. also thanks  for the
 support by  the National Natural  Science Foundation of  China (Grant
 No. J1210039). A.L. is supported in  part by NSF AST-1311804 and NASA
 NNX14AF68G.  This  research has made  extensive use of  the NASA/IPAC
 Extragalactic Database (NED) which is  operated by the Jet Propulsion
 Laboratory, California  Institute of Technology, under  contract with
 the  National Aeronautics  and  Space Administration.   This work  is
 based  [in  part]  on  observations   made  with  the  Spitzer  Space
 Telescope,  which  is  operated  by the  Jet  Propulsion  Laboratory,
 California Institute of Technology under  a contract with NASA."  The
 {\it Galaxy  Evolution Explorer}  (GALEX) is  a NASA  Small Explorer,
 launched  in   April  2003.    We  acknowledge  NASA's   support  for
 construction, operation, and science analysis for the GALEX mission.












\bsp	
\label{lastpage}
\end{document}